\documentclass[final,3p,times,twocolumn]{elsarticle}
\biboptions{comma,sort&compress}
\usepackage{ecrc}
\usepackage{here}
\usepackage{graphicx}
\usepackage{epsfig}
\usepackage{epstopdf}
\usepackage{amsmath}

\def\nin{\noindent}
\def\beq{\begin{equation}}
\def\eeq{\end{equation}}
\def\bea{\begin{eqnarray}}
\def\eea{\end{eqnarray}}
\def\nnb{\nonumber}
\def\la{\langle}
\def\ra{\rangle}

\usepackage{graphicx}
\usepackage{here}
\def\beq{\begin{equation}}
\def\eeq{\end{equation}}
\def\bea{\begin{eqnarray}}
\def\eea{\end{eqnarray}}
\def\bq{\begin{quote}}
\def\eq{\end{quote}}

\def\nnb{\nonumber}

\def\nnb{\nonumber}
\def\la{\langle}
\def\ra{\rangle}
\def\nin{\noindent}
\def\ba{\begin{array}}
\def\ea{\end{array}}

\def\als{\alpha_s}

\def\gg2{ \la\alpha_s G^2 \ra}
\def\gg3{g^3f_{abc}\la G^aG^bG^c \ra}
\def\ggg4{\la\als^2G^4\ra}

\def\beq{\begin{equation}}
\def\enq{\end{equation}}
\def\beqa{\begin{eqnarray}}
\def\enqa{\end{eqnarray}}
\def\nnb{\nonumber}

\newcommand{\rag}{\rangle}
\newcommand{\lag}{\langle}

\def\ln{\mbox{Log}}
\def\gg{\lag g^{2}_{s} G^2 \rag}
\def\ggg{\lag g^{3}_{s}G^3\rag}

\volume{00}
\firstpage{1}
\journalname{Nuclear and Particle Physics Proceedings }
\runauth{D. Rabetiarivony}
\jnltitlelogo{Nuclear and Particle Physics Proceedings }

\begin{document}
\begin{frontmatter}

\title{$DK$ and $BK$-like spectra from Laplace sum rule at NLO \tnoteref{text1}}


\author[label2,label3]{S. Narison\fnref{fn0}}
\fntext[fn0]{ICTP-Trieste consultant for Madagascar.}
\ead{snarison@yahoo.fr}
\address[label2]{Laboratoire Univers et Particules de Montpellier, CNRS-IN2P3, Case 070, Place Eug\`ene Bataillon, 34095 - Montpellier, France.}
\address[label3]{Institute of High Energy Physics of Madagascar (iHEPMAD), University of Ankatso, Antananarivo 101, Madagascar.}

\author[label3]{A. Rabemananajara\fnref{fn1}}
\fntext[fn1]{Speaker}
\ead{achrisrab@gmail.com}

\author[label3]{D. Rabetiarivony}

\ead{rd.bidds@gmail.com}

\tnotetext[text1]{Talk given at QCD21 International Conference (05--09 July 2021, Montpellier--FR)}

\pagestyle{myheadings}
\markright{ }

\begin{abstract}
\noindent
Encouraged  by the agreement, with the recent LHCb data on the $D^-K^+$ invariant mass from $B \rightarrow D^+D^-K^+$ decay, of our results for the masses of the $0^+$ and $1^-$ open charm $(\bar c\bar d)(u s)$ tetraquarks and molecules states from QCD spectral sum rules within stability criteria, which we review here, we extend our analysis to the $b$-quark channel.
We find, in the $0^+$ case the lowest mass $M_{BK}=5195(15)$ MeV with $f_{BK}=8.3(2.4)$ keV and three (almost) degenerate states  having respectively the masses $M_{SS}=5702(60)~{\rm MeV}$, $M_{AA}=5661(75)~{\rm MeV}$ and $M_{B^*K^*}=5720(71)~{\rm MeV}$ and couplings $f_{SS}=22.2(2.3)~{\rm keV}$, $f_{AA}=30.1(3.1)~{\rm keV}$ and $f_{B^*K^*}=26.5(2.8)~{\rm keV}$, from which we can associate a scalar tetramole with  $M_{\mathcal T_{\mathcal M_0}}=5694(69)~{\rm MeV}~\text{and} ~ f_{\mathcal T_{\mathcal M_0}}=26.5(2.7)~{\rm keV}$. In the spin 1 case, we find four (almost) degenerate states associated with a tetramole having $M_{\mathcal T_{\mathcal M_1}}=5700(81)~{\rm MeV}~\text{and} ~ f_{\mathcal T_{\mathcal M_1}}=16.2(2.6)~{\rm keV}$. For the first radial excitation of the $BK$ molecule, we have $M_{(BK)_1} = 6265(146)$ MeV and  $f_{(BK)_1} = 22.8(3.2)$ keV. For the remaining states, we associate a scalar and vector tetramoles having respectively $M_{\mathcal (T_{\mathcal M_0})_1}=7439(314)~{\rm MeV}, ~ f_{\mathcal (T_{\mathcal M_0})_1}=74.7(8.4)~{\rm keV}$ and
$M_{\mathcal (T_{\mathcal M_1})_1}=7544(345)~{\rm MeV}, ~ f_{\mathcal (T_{\mathcal M_1})_1}=33.0(6.7)~{\rm keV}$. 
\end{abstract} 
\scriptsize
\end{frontmatter}


\vspace*{-1.2cm}
\section{Introduction}
\vspace*{-0.2cm}
\nin
 We have systematically calculated the masses and couplings of some possible configurations of the $0^+$ and $1^-$ molecules and tetraquarks open charm $(\bar c\bar d)(u s)$ tetraquarks and molecules states from  the inverse Laplace Transform (LSR) \cite{LP1,LP2,LP3,LP4,LP5} of QCD spectral sum rules (QSSR) \footnote{For reviews, see \cite{SR1,SR2,SR3,SR4,SR5,SR6,SR7,SR8,SR9,SR10,JAMI2b}}within stability criteria in Ref.\cite{DK} which we have used to explain the recent LHCb data on the $D^+K^-$ invariant mass \cite{LHCb1, LHCb2}.


We concluded that the $2400$ MeV bump around $DK$ threshold can be due to $DK$ scattering amplitude $\oplus$ the $DK(2400)$ lowest mass molecule. The ($0^+$) $X_0(2866)$ and $X_J(3150)$ (if it is a $0^+$ state) can e.g result from a mixing of the Tetramole with the $1^\text{st}$ radial excitation $(DK)_1$ of the molecule state ($DK$) with a tiny mixing angle $\theta_0 = (5.2 \pm 1.9)^\circ$. The $(1^-)$ $X_1(2904)$ and $X_J(3350)$ (if it is a $1^-$ state) can result from a mixing of the Tetramole with its $1^\text{st}$ radial excitation with a tiny mixing angle $\theta_1 = (9.1 \pm 0.6)^\circ$. In this work, based on these ~coherent results, we extend our work to the b-channel, to estimate, from LSR, the masses and couplings of some $BK$-like spectra  at NLO in the PT-series. 
\vspace*{-0.2cm}
\section{The Laplace sum rule (LSR)}
We shall work with the Finite Energy version of the QCD Inverse Laplace sum rules and their ratios:
\bea
\mathcal L_n^c(\tau ,\mu )&=&\int _{\left(M_b+m_s\right){}^2}^{t_c} {dt} \ t^n\ e^{-t \tau }\frac{1}{\pi }\text{Im} \Pi _{\mathcal M,\mathcal T}(t,\mu )\nnb\\
\mathcal R_n^c(\tau )&=&\frac{\mathcal L_{n+1}^c}{\mathcal L_n^c}
\eea
where $M_b$ and $m_s$ are the on-shell / pole bottom and
running strange quark masses, $\tau$ is the LSR variable,
$n = 0 ; 1$ is the degree of moments, $t_c$ is the threshold of the “QCD continuum” which parametrizes, from the discontinuity of the Feynman diagrams, the spectral function evaluated by the calculation of the correlator defined as:
{\footnotesize
\begin{flalign}
\Pi _{\mathcal M,\mathcal T}(t,\mu )(q^2)=i\int d^4x\ e^{-iqx}\langle 0|T[\mathcal O^J_{\mathcal M,\mathcal T} (x)\mathcal O^{J\ \dagger}_{\mathcal M,\mathcal T}(0)]|0\rangle
\end{flalign}
}
where $O^J_{\mathcal M,\mathcal T} (x)$ are the interpolating currents for the
tetraquarks $\mathcal T$ and molecules $\mathcal M$ states. The superscript $J$ refers to the spin of the particles.
\begin{itemize}
\item $\tau$ is an arbitrary parameter, then the mass and coupling have to be independent on him. We must get a stability w.r.t $\tau$.
\item We find the beginning of the stability for different values of $t_c$.
\item The value of the mass (or coupling) was taken at the minimum or the inflexion point.
\end{itemize}
\section{Interpolating currents}
The interpolating currents for the molecules and for tetraquark states are given in Tables \ref{currents}.
\begin{footnotesize}
\begin{center}
\begin{table}[h]
\caption{Interpolating operators describing the scalar ($0^+$) and vector ($1^-$) molecules and tetraquark states used in this work}
\begin{tabular}{ll}
\hline
\hline
 &Tetraquarks  \\ 

$0^+$ & $\mathcal O_{SS}=\epsilon _{{ijk}} \epsilon _{{mnk}}\left(u_i^TC \gamma _5d_j\right)\left(\bar{b}_m\gamma _5C \bar{s}_n^T\right)$  \\ 

 & $\mathcal O_{AA}=\epsilon _{{ijk}} \epsilon _{{mnk}}\left(u_i^TC \gamma _{\mu }d_j\right)\left(\bar{b}_m\gamma ^{\mu }C \bar{s}_n^T\right)$  \\ 
 
$1^-$ & $\mathcal O_{SV}=\epsilon _{{ijk}} \epsilon _{{mnk}}\left(u_i^TC \gamma _5d_j\right)\left(\bar{b}_m\gamma _{\mu }\gamma _5C \bar{s}_n^T\right)$  \\ 
 
 & $\mathcal O_{PA}=\epsilon _{{mnk}}\epsilon _{{ijk}}\left(\bar{b}_mC \bar{s}_n^T\right)\left(u_i^TC \gamma _{\mu }d_j\right)$  \\ 
\hline 
 &Molecules\\
 
 $0^+$ & $\mathcal O_{B^*K^*}=\left(\bar{b} \gamma ^{\mu }d\right)\left(\bar{s}\gamma _{\mu } u\right)$ \\ 

 &  $\mathcal O_{B K}=\left(\bar{b}d\right)\left(\bar s u\right)$ \\ 
 
$1^-$  & $\mathcal O_{B_1K}=\left(\bar{b} \gamma _{\mu } \gamma _5 d\right)\left(\bar{s} \gamma _5 u\right)$ \\ 
 
& $\mathcal O_{B_0^*K^*}=\left(\bar{b}d\right)\left(\bar s \gamma _{\mu }u\right)$ \\
\hline
\hline  
\end{tabular} 

\label{currents}
\end{table}
\end{center}
\end{footnotesize}
\subsection*{$\bullet$ Extracting the lowest ground state mass and coupling}
In \cite{HEP3}, we have extracted the lowest ground state mass by using the minimal duality ansatz (MDA):
\begin{flalign}
\frac{1}{\pi}\text{Im} \Pi _{\mathcal M,\mathcal T}\simeq f_{\mathcal M,\mathcal T}^2 M_{\mathcal M,\mathcal T}^8 \delta(t-M_{\mathcal M,\mathcal T}^2)+\Theta(t-t_c)
\end{flalign}
The decay constant $f_M$ (analogue to $f_\pi=132$ MeV) for the molecule state is defined as: 
\bea
\langle 0|\mathcal O_{\bar BK}|{\bar BK}\rangle&=&f_{\bar BK} M^4_{\bar BK}\nnb\\
\langle 0|\mathcal O^\mu_{\bar B^*K}|{\bar B^*K}\rangle&=&\epsilon^\mu f_{\bar B^*K} M^5_{\bar B^*K}
\eea
and analogously for the one $f_{\mathcal T}$ of the tetraquark state. Interpolating currents constructed from bilinear (pseudo)scalar currents are not renormalization group invariants such that the corresponding decay constants possess anomalous dimension:
\bea
f_{\bar{B}K}(\mu )&=&\hat{f}_{\bar{B}K}\left(-\beta _1a_s\right){}^{4\left/\beta _1\right.}\left(1-k_f \ a_s\right)\nnb\\
f_{\bar{B}^*K}(\mu )&=&\hat{f}_{\bar{B}^*K}\left(-\beta _1a_s\right){}^{2\left/\beta _1\right.}\left(1-k_f\left.a_s\right/2\right)
\eea
where: $\hat f_M$ is the renormalization group invariant coupling and $-\beta_1 = (1/2)(11- 2n_f /3)$ is the first coefficient of the QCD $\beta$-function for $n_f$ flavours. $a_s \equiv (\alpha_s/\pi)$ is the QCD coupling and $k_f = 2.028(2.352)$ for $n_f = 4(5)$ flavours. Within a such parametrization, one obtains:
\bea
\mathcal R_0^c \equiv \mathcal{R} \simeq M^2_\mathcal{M,T}
\eea
where $M_\mathcal{M,T}$ is the lowest ground state molecule or tetraquark mass. 
 \subsection*{$\bullet$ Higher orders PT corrections to the spectral functions}
 We extract the NLO PT corrections by considering that the molecule/tetraquark two-point
spectral function is the convolution of the two ones built from two quark bilinear currents (factorization). In this way, we obtain \cite{CONV,SNPIVO}:
\bea
&&\frac{1}{\pi }\text{Im} \Pi _{\mathcal M,\mathcal T}(t)=\theta \left(t-\left(M_b+m_s+m_d\right){}^2\right)\left(\frac{k}{4\pi }\right)^2t^2\nnb\\
&&\times  \underset{\left(M_b+m_d\right){}^2}{\overset{\left(\sqrt{t}-m_s\right){}^2}{\int }}\text{dt}_1\underset{m_s^2}{\overset{\left(\sqrt{t}-\sqrt{t_1}\right){}^2}{\int }}\text{dt}_2\lambda ^{1/2}\mathcal K^H
\label{eq1}
\eea
where $k$ is an appropriate normalization factor, $M_b$ is the on-shell/pole perturbative heavy quark mass.
\begin{flalign}
&\mathcal K^{\text{SS},\text{PP}}\equiv  \left(\frac{t_1}{t}+\frac{t_2}{t}-1\right){}^2\times \frac{1}{\pi }\text{Im} \psi ^{S,P}\left(t_1\right)\frac{1}{\pi }\text{Im} \psi ^{S,P}\left(t_2\right)\nnb\\
&\mathcal K^{\text{VV},\text{AA}}\equiv \left[\left(\frac{t_1}{t}+\frac{t_2}{t}-1\right){}^2+8 \left(\frac{t_1t_2}{t^2}\right)\right]\nnb\\&\ \ \ ~~~~~~~~\times \frac{1}{\pi }\text{Im} \psi ^{V,A}\left(t_1\right)\frac{1}{\pi }\text{Im} \psi ^{V,A}\left(t_2\right)
\end{flalign}
for spin zero scalar state and:
\begin{flalign}
\mathcal K^{\text{VS},\text{AP}}\equiv 2\lambda  \times \frac{1}{\pi }\text{Im} \psi ^{V,A}\left(t_1\right)\frac{1}{\pi }\text{Im} \psi ^{S,P}\left(t_2\right)
\end{flalign}
for spin one vector state, with the phase space factor:
\begin{flalign}
\lambda =\left(1-\frac{\left(\sqrt{t_1}-\sqrt{t_2}\right){}^2}{t}\right)\left(1-\frac{\left(\sqrt{t_1}+\sqrt{t_2}\right){}^2}{t}\right)
\end{flalign}
For tetraquark, one interchanges $s$ and $d$ in the integrals of Eq.(\ref{eq1}). We have taken $m_u = 0$ for simplifying the expression but we shall also neglect $m_d$ in the numerical analysis.
\subsection*{$\bullet$ QCD input parameters}
The PT QCD parameters which appear in this analysis are $\alpha_s$,  strange and bottom quark masses $m_{s,b}$ (the light quark masses have been neglected). 
We also consider non-perturbative condensates  which are the quark condensate $\langle\bar q q\rangle$, the two-gluon condensate $\langle g^2 G^2\rangle$, the mixed condensate $\langle g\bar q G q\rangle$, the three-gluon condensate $\langle g^3 G^3\rangle$, and  the four-quark condensate $\rho\langle\bar q q\rangle^2$,  where $\rho\simeq (3 \sim 4)$ indicates the deviation from the four-quark vacuum saturation. Their values are given in Table \ref{tab:parameter}.

{\scriptsize
\begin{table}[h]
\setlength{\tabcolsep}{.2pc}
 \caption{QCD input parameters estimated from QSSR (Moments, LSR and ratios of sum rules).}  
    {\footnotesize
 {\begin{tabular}{@{}lll@{}}
&\\
\hline
\hline
Parameters&Values& Ref.    \\
\hline
$\alpha_s(M_Z)$& $0.1181(16)(3)$&\cite{SNP1,SNP2}\\
$\hat m_s$&$(0.114\pm0.006)$ GeV &\cite{SR4,SMS1,SMS2}\\
$\overline{m}_b(m_b)$&$4202(8)$ MeV& \cite{SNP3}\\
$\hat \mu_q$&$(253\pm 6)$ MeV&\cite{SR4,SMS1,SMS2}\\
$M_0^2$&$(0.8 \pm 0.2)$ GeV$^2$&\cite{SR4,IOFFEp,JAMI2a,JAMI2b,JAMI2c,HEIDb,HEIDc,SNhl}\\
$\la\alpha_s G^2\ra$& $(6.35\pm 0.35)\times 10^{-2}$ GeV$^4$&
\cite{SNP1}\\
$\la g^3  G^3\ra$& $(8.2\pm 2.0)$ GeV$^2\times\la\alpha_s G^2\ra$&
\cite{SNH10a,SNH10b,SNH10c}\\
$\rho \alpha_s\la \bar qq\ra^2$&$(5.8\pm 0.9)\times 10^{-4}$ GeV$^6$&\cite{G3,SNTAU,LNT,JAMI2a,JAMI2b,JAMI2c}\\
\hline\hline
\end{tabular}}
}
\label{tab:parameter}
\end{table}
} 
The Renormalization Group Invariant parameters are defined as \cite{SR3,SR4}:
\begin{flalign}
&\bar{m}_s(\tau )=\hat{m}_s\left(-\beta_1 a_s\right){}^{-2/\beta_1}, ~~~~~~\langle\bar{q}q\rangle(\tau )=-\hat{\mu }_q^3\left(-\beta _1a_s\right)^{2/\beta_1},\nnb\\
&\langle\bar{q}Gq\rangle(\tau )=-M_0^2\hat{\mu }_q^3\left(-\beta _1a_s\right){}^{1/3\beta _1}
\end{flalign}
$\hat{\mu }_q$ is the spontaneous RGI light quark condensate \cite{FNR}. The running bottom mass $m_b$ is related to the on-shell (pole) mass $M_b$ used to compute the two-point correlator from the NLO relation \cite{SPEC1,TAR,COQ,SNmass98b,BIN}:
\begin{flalign}
\scriptsize
M_b(\mu )=\bar{m}_b(\mu )\left[1+\frac{4}{3}a_s(\mu )+\ln \left(\frac{\mu }{M_b}\right){}^2a_s(\mu )+\mathcal O\left(a_s^2\right)\right]\nnb\\
\end{flalign}
\section{Molecules and tetraquarks}
We will study the mass and coupling of some $BK$-like states. As the analysis will be performed using the same techniques, we shall illustrate the case of the ${SS} ~(0^+)$ tetraquark.
\subsection*{$\bullet$ $f_{SS}$ and $M_{SS}$}
\begin{center}
\begin{figure}[h]
\includegraphics[width=3.8cm, height=3.1cm]{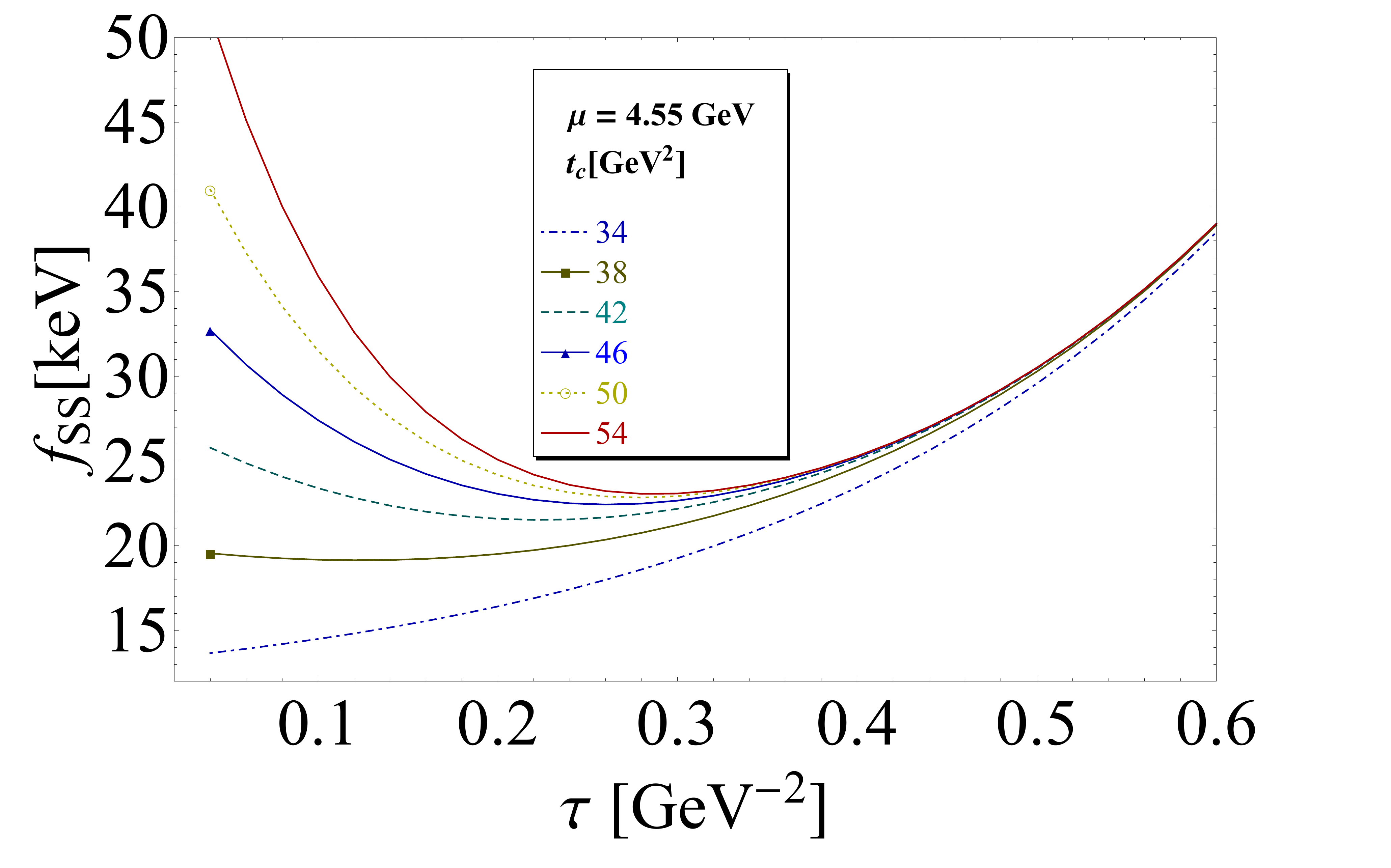}
\includegraphics[width=3.8cm, height=3cm]{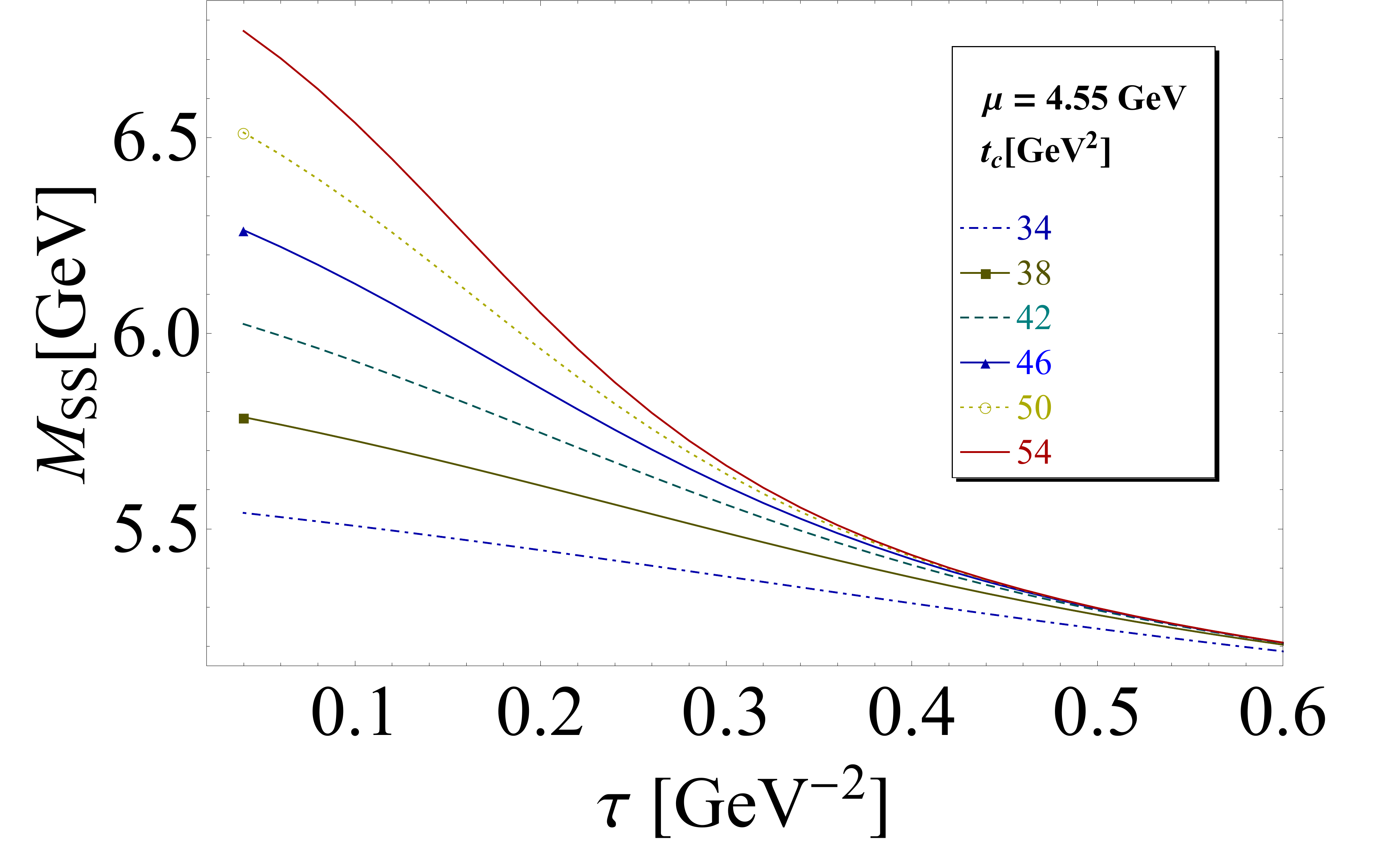}
\caption{{\footnotesize $f_{SS}$ and $M_{SS}$ at NLO  as function of $\tau$ for different values of $t_c$, for $\mu=4.55 \rm GeV$}}
\label{tau}
\end{figure}
\end{center}

We study the behaviour of the coupling and mass in term of the LSR variable $\tau$ for different values of $t_c$ at NLO as shown in Fig \ref{tau}.  We consider as  result the one corresponding to the beginning of the $\tau$-stability for $t_c$ =42 GeV$^2$ and $\tau\simeq$ 0.22 GeV$^{-2}$ until the one where $t_c$-stability is reached for $t_c\simeq$ 50 GeV$^2$ and $\tau\simeq$ 0.28 GeV$^{-2}$.
\subsection*{$\bullet$ $\mu$-stability}

\begin{center}
\begin{figure}[h]
\includegraphics[width=3.8cm, height=3cm]{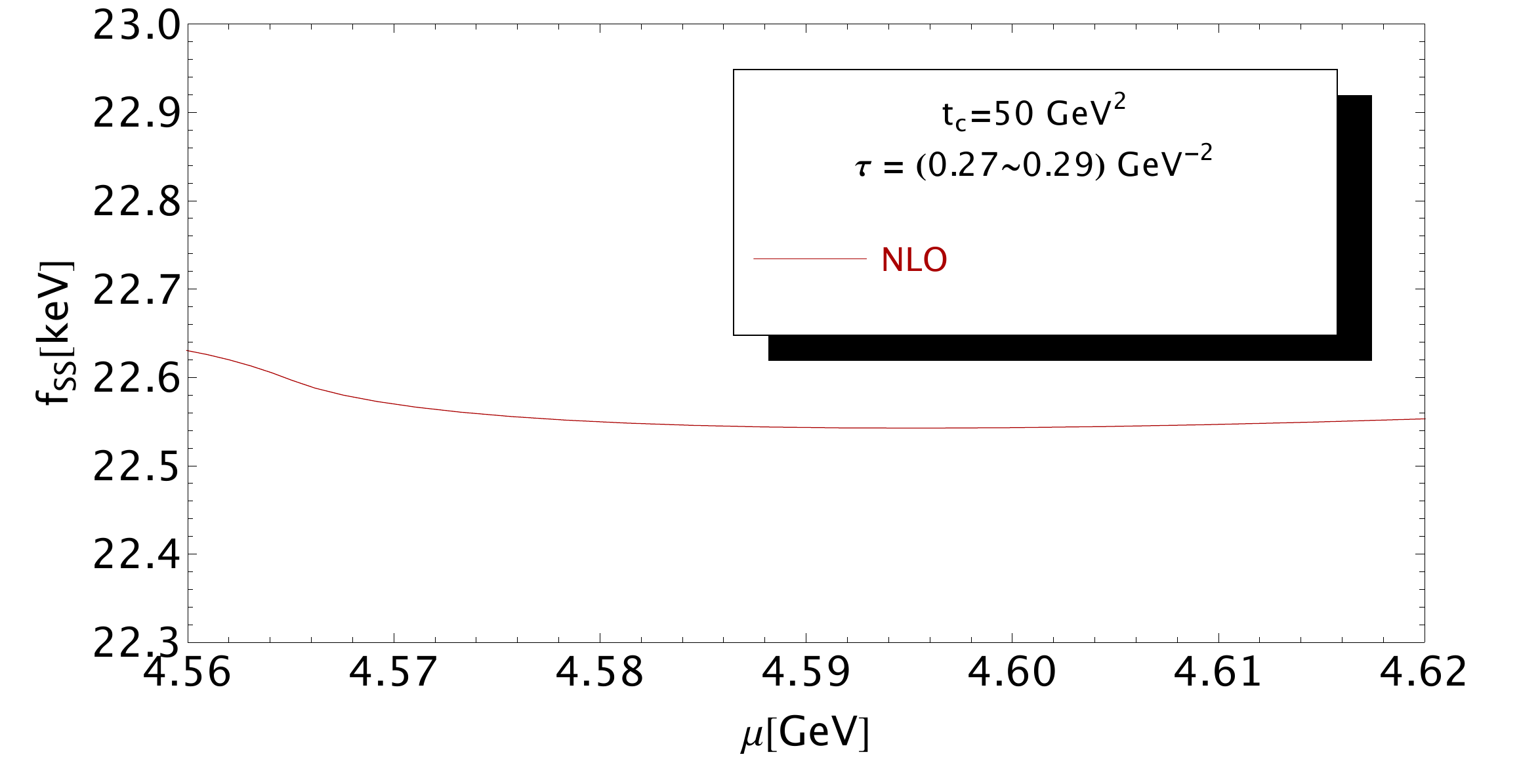}
\includegraphics[width=3.8cm, height=3.05cm]{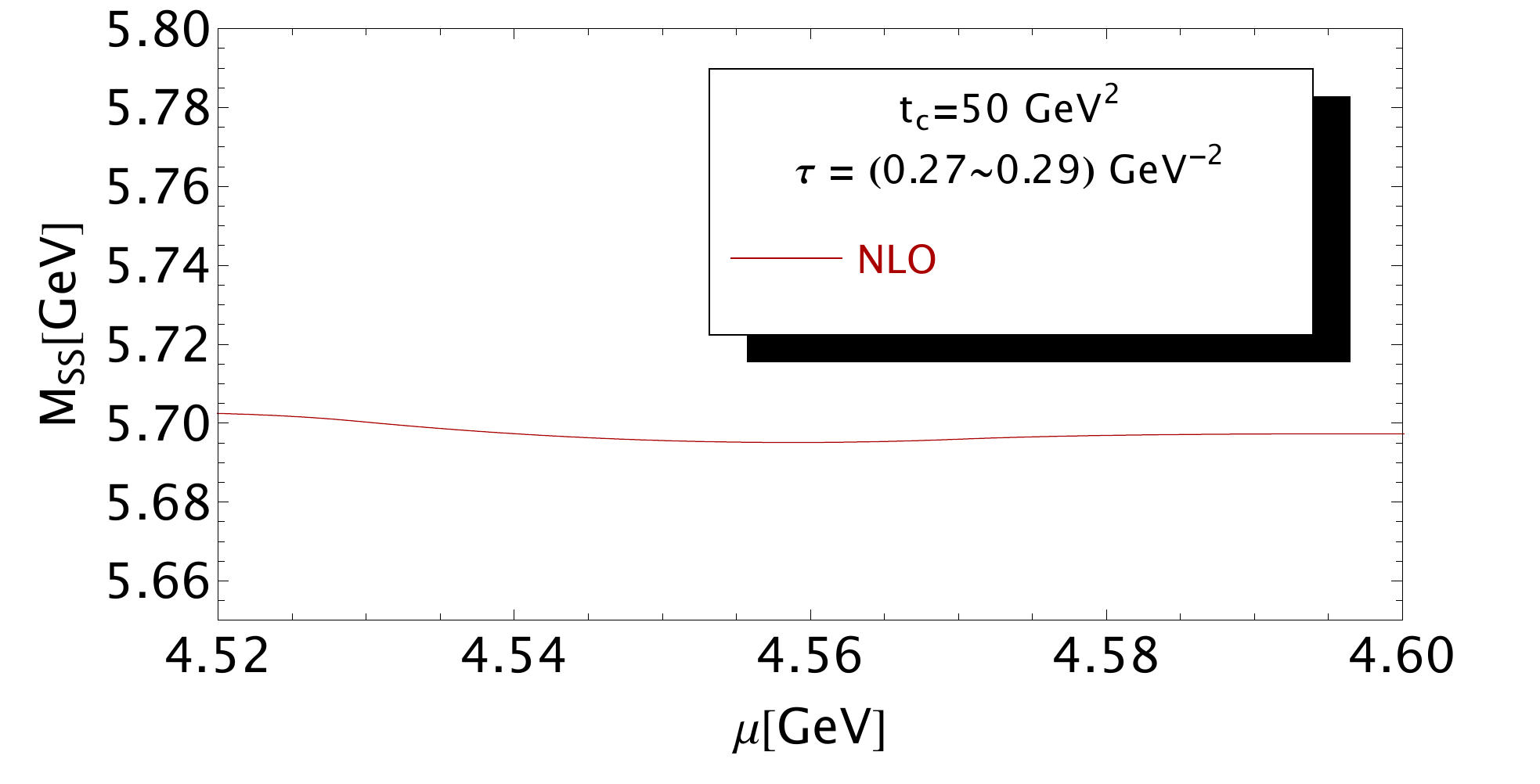}
\caption{{\footnotesize $f_{SS}$ and $M_{SS}$ at NLO  as function of $\mu$ for $t_c= 50 ~\ \rm GeV^2$ in the region of $\tau$ stability }}
\label{mustab}
\end{figure}
\end{center}

In Fig.\ref{mustab}, we show the $\mu$-dependence of the results for given $t_c=50$ GeV$^{-2}$ and $\tau  \sim 0.29$ GeV$^{-2}$. One finds a common stability for $\mu = (4.55 \pm 0.25) $GeV.
\subsection*{$\bullet$ LO versus NLO results}
\begin{center}
\begin{figure}[h]
\includegraphics[width=3.8cm, height=3cm]{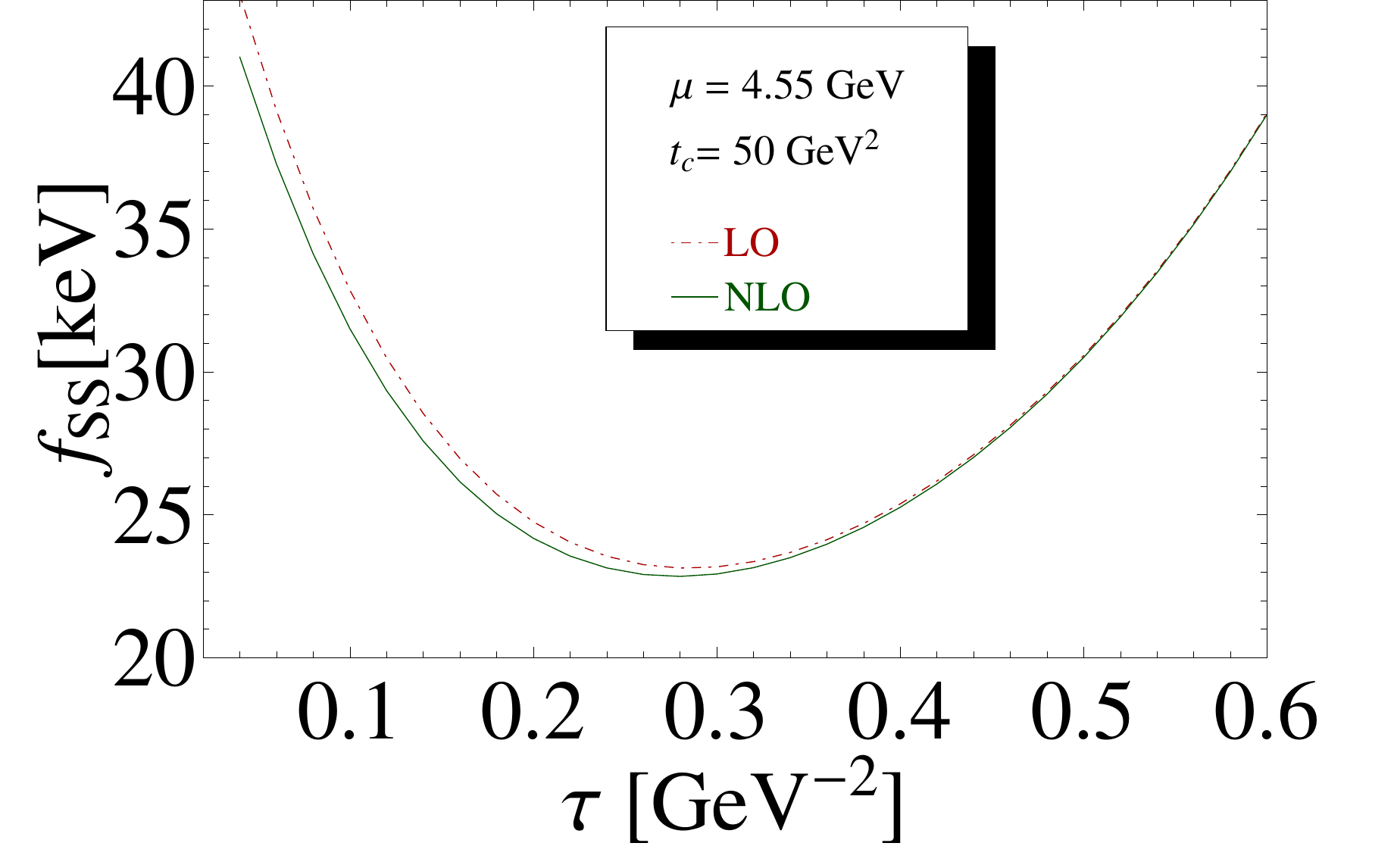}
\includegraphics[width=3.8cm, height=3cm]{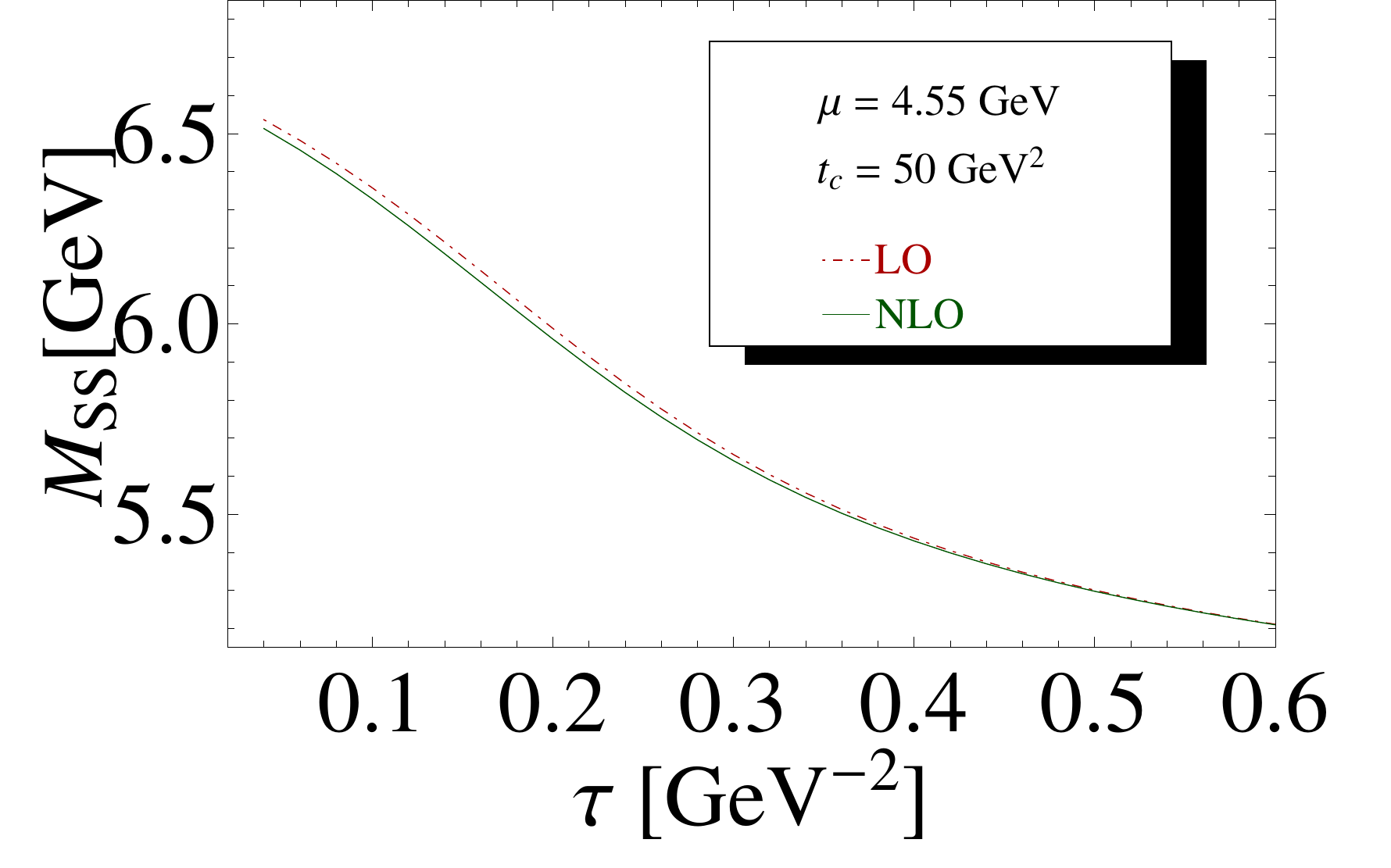}
\caption{{\footnotesize Comparison of $f_{SS}$ and $M_{SS}$ at LO and NLO  as function of $\tau$ for $t_c= 50 \rm GeV^2$ and $\mu = 4.55 \rm GeV$}}
\label{lonlo}
\end{figure}
\end{center}

We compare in Fig.\ref{lonlo} the $\tau$-behaviour of the mass and coupling for fixed $t_c= 50 ~\ \rm GeV^2$ and $\mu = 4.55 $GeV at LO and NLO of perturbative QCD in the MS-scheme. One can notice that the $\alpha_s$ corrections are small for $SS$ which, at the stability point, decrease the mass by 18 MeV and  the coupling by 0.29 keV. According to these analysis, the final results are compiled in Tables \ref{results}.
\section{Tetramoles}
\label{threshold}
Our results indicate that the molecules and tetraquark states leading to the same final states are almost degenerated in masses. Therefore, according to conclusion in \cite{DK}, we expect that the "physical state" is a combination of almost degenerated molecules and tetraquark states with the same quantum, numbers $J^{PC}$ which we shall call: Tetramole ($\mathcal T_{\mathcal M_J}$).
\subsection*{$\bullet$ Tetramole $0^+$}
Taking  our results in Table \ref{results}, one can see that we have three (almost) degenerate states:
\bea
&M_{SS}=5702(60)~{\rm MeV},~ M_{AA}=5661(75)~{\rm MeV}\nnb\\
&\text{and}~~~M_{B^*K^*}=5720(71)~{\rm MeV}\nnb
\eea
and their couplings to the corresponding currents are almost the same:
\bea
&f_{SS}=22.2(2.3)~{\rm keV},~ f_{AA}=30.1(3.1)~{\rm keV}\nnb\\
&\text{and}~~~f_{B^*K^*}=26.5(2.8)~{\rm keV}\nnb
\eea
We assume that the physical state, hereafter called Tetramole ($\mathcal T_{\mathcal M_J}$), is a superposition of these nearly degenerated hypothetical states having the same quantum
numbers. Taking its mass and coupling as (quadratic) means of the previous numbers, we obtain:
\begin{align}
M_{\mathcal T_{\mathcal M_0}}=5694(69)~{\rm MeV}~\text{and} ~ f_{\mathcal T_{\mathcal M_0}}=26.5(2.7)~{\rm keV}\nnb
\end{align}
\subsection*{$\bullet$ Tetramole $1^-$}
In Table \ref{results}, in the case of spin 1, we have four degenerate states with the same masses around 5700 GeV, taking the (quadratic) means, we obtain:
\bea
M_{\mathcal T_{\mathcal M_1}}=5700(81)~{\rm MeV}~\text{and} ~ f_{\mathcal T_{\mathcal M_1}}=16.2(2.6)~{\rm keV}\nnb
\eea

\section{The first radial excitation}
According to \cite{DK}, we also extend our analysis by using a "Two resonances" + $\theta(t-t_c)$ "QCD continuum" parametrization of the spectral function, we illustrate again the case of $SS$ tetraquark. To enhance the contribution of the first radial excitation, we will also work with the ratio of moments $\mathcal{R}_1$ for getting the mass of  $(SS)_1$.

\subsection*{$\bullet$ $\tau$ and $t_c$-stability}
\begin{center}
	\begin{figure}[h]
		\includegraphics[width=3.8cm, height=3cm]{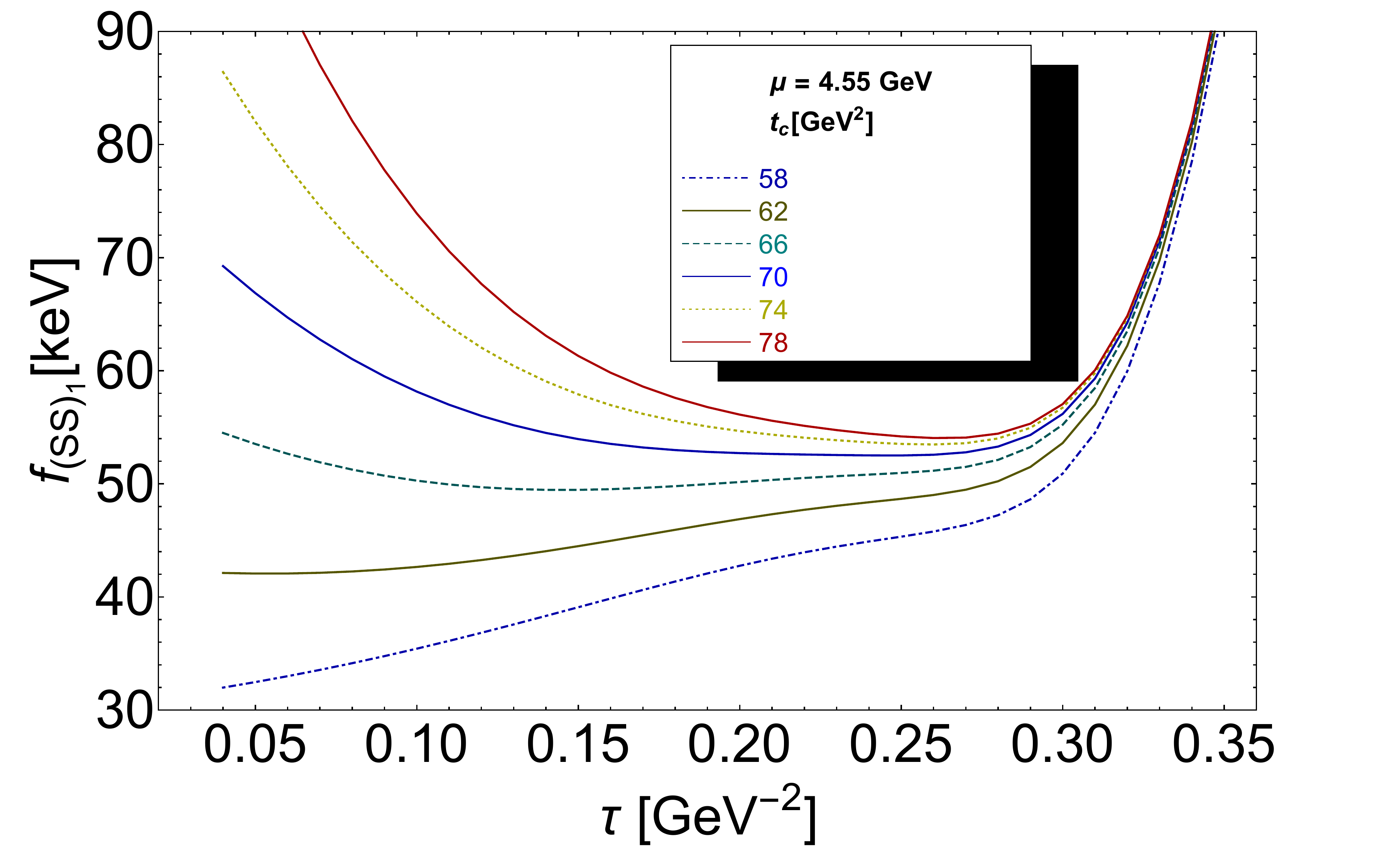}
		\includegraphics[width=3.8cm, height=3cm]{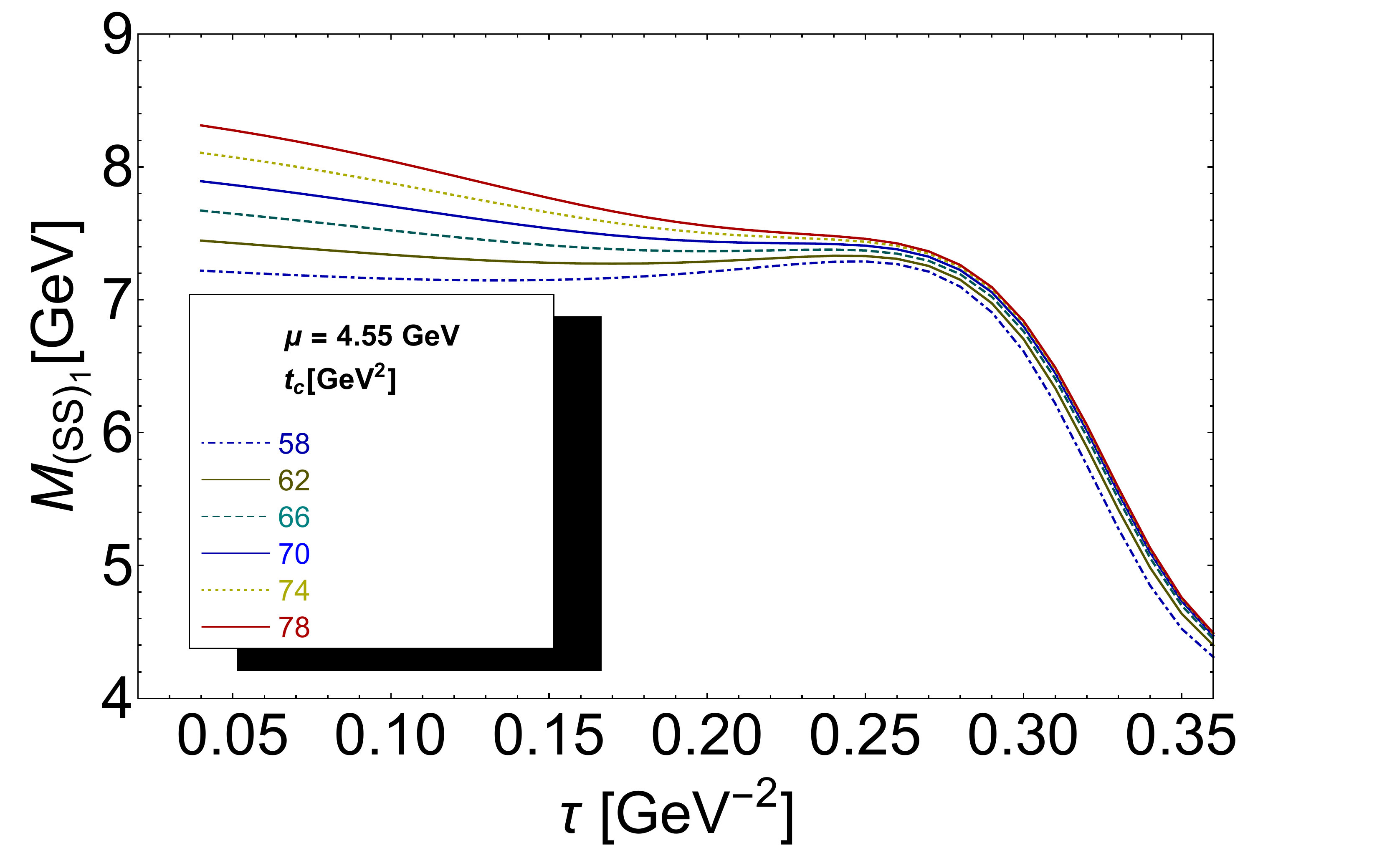}
		\caption{{\footnotesize $f_{(SS)_1}$ and $M_{(SS)_1}$ at NLO  as function of $\tau$ for different values of $t_c$, for $\mu=4.55 \rm GeV$}}
		\label{tau1}
	\end{figure}
\end{center}

We show in Fig. \ref{tau1} the $\tau$ and $t_c$-behaviours of the coupling from $\mathcal{L}_0^c$ and  the ones of the mass from  $\mathcal{R}_1$ using as input the values of the lowest ground state mass and coupling.  One can notice that the coupling from $\mathcal{L}_0^c$ stabilizes for $\tau \simeq (0.15 \sim  0.26)$ GeV$^{-2}$ which is slightly lower than the value $\tau = 0.29$ GeV$^{-2}$ corresponding to the one-resonance parametrization. The  values of $t_c$ are 66 to 74 GeV$^{-2}$ compared to 42 to 50 GeV$^{-2}$ for the one resonance case. The result is given in Table \ref{results1}.

\subsection*{$\bullet$ Tetramole $0^+$}
One can also notice from Tables \ref{tab:res1} that the radial excitations other than the one of $BK$ with a mass of 6265(146) MeV are almost degenerated around 7.4 GeV from which one can extract the masses and couplings (geometric mean) of the spin 0 excluding $(BK)_1$. \\
Then, for the $0^+$ case, we have three degenerate states 
having the mass and coupling (quadratic means):
\begin{align}
M_{\mathcal (T_{\mathcal M_0})_1}=7439(314)~{\rm MeV}~\text{and} ~ f_{\mathcal (T_{\mathcal M_0})_1}=74.7(8.4)~{\rm keV}\nnb
\end{align}
\subsection*{$\bullet$ Tetramole $1^-$}
For the spin $1^-$ case, in Table \ref{tab:res1}, we have four degenerate states around 7.5 GeV.
 Taking the (quadratic) means, we  have: 
\begin{align}
 M_{\mathcal (T_{\mathcal M_1})_1}=7544(345)~{\rm MeV}~\text{and} ~ f_{\mathcal (T_{\mathcal M_1})_1}=33.0(6.7)~{\rm keV}\nnb
\end{align} 
 
\section{Conclusions}
\nin
$\bullet$ We have presented improved predictions of QSSR for the masses and couplings of the $0^+$ and $1^-$ $BK$-like molecule and four-quark states at NLO  of PT series and including up to dimension six non-perturbative condensates.\\  
$\bullet$ Our analysis has been done within stability criteria with respect to the LSR variable $\tau$, the QCD continuum threshold $t_c$ and the subtraction constant $\mu$ which have provided successful predictions in different hadronic channels. The optimal values of the masses and couplings have been extracted at the same value of these parameters where the stability appears as an extremum and/or inflection points.\\
$\bullet$ The results for all the 4-quarks and molecule currents discussed here are below their thresholds. \\
$\bullet$ NLO radiative corrections are essential for giving a meaning on the input value of the charm and bottom quark masses which plays an important role in the analysis. We consider our results as improvement and a completion of the results obtained to LO from QCD spectral sum rules.\\
$\bullet$ We find as lowest mass $M_{BK}=5195(15)$ MeV with $f_{BK}=8.3(2.4)$ keV. We also obtain, in the $0^+$ case,  three (almost) degenerate states, to which we can associate a scalar tetramole with  $M_{\mathcal T_{\mathcal M_0}}=5694(69)~{\rm MeV}~\text{and} ~ f_{\mathcal T_{\mathcal M_0}}=26.5(2.7)~{\rm keV}$. In the spin 1 case, we find four degenerates states leading to a tetramole having a mass $M_{\mathcal T_{\mathcal M_1}}=5700(81)~{\rm MeV}~\text{and}$ and a coupling $f_{\mathcal T_{\mathcal M_1}}=16.2(2.6)~{\rm keV}$. \\
$\bullet$ In an analogous way, we predict the first radial excitations of these states which give: $M_{(BK)_1} = 6265(146)$ MeV and  $f_{(BK)_1} = 22.8(3.2)$ keV. We associate for the other states a scalar and vector tetramoles having respectively $M_{\mathcal (T_{\mathcal M_0})_1}=7439(314)~{\rm MeV}, ~ f_{\mathcal (T_{\mathcal M_0})_1}=74.7(8.4)~{\rm keV}$ and $M_{\mathcal (T_{\mathcal M_1})_1}=7544(345)~{\rm MeV}, ~ f_{\mathcal (T_{\mathcal M_1})_1}=33.0(6.7)~{\rm keV}$\\  
$\bullet$ Our approach doesn't discern, within the errors, the molecules and tetraquarks states.  

\begin{table*}[hbt]
\setlength{\tabcolsep}{1.5pc}
\caption{LSR predictions, at NLO, for the decay constants and 
masses of the ground state ($f_0$, $M_0$), 
for the molecules, tetraquark  and the predicted tetramole states. The symbol ``$\ast$'' indicates that $BK$ did not contribute to the tetramole results}
\label{tab:res}
\begin{tabular*}{\textwidth}{@{}l@{\extracolsep{\fill}}
cccc}
\hline
\hline
Observables\, & $M_0$ (MeV) & $f_0$ (keV) 
& $M_{\mathcal T_{\mathcal M_{0,1}}}$ (MeV) & $M_{\mathcal T_{\mathcal M_{0,1}}}$ (keV)\\
\hline\\
{\bf \boldmath $0^+$ States}\\
\cline{0-0}
{\it Molecule}\\
${BK}$ & 5195(15) & 08.3(2.4) &  $\ast$ &  $\ast$ \\
${B^*K^*}$ & 5720(71) & 26.5(2.8) & & \\
 &  &  &5694(69) & 26.5(2.7)\\
%
{\it Tetraquark}\\
${SS}$ & 5702(60) & 22.2(2.3) &  & \\
${AA}$ & 5661(75) & 30.1(3.1) &  & \\\\
%
{\bf \boldmath $1^-$ States}\\
\cline{0-0}
{\it Molecule}\\
${B_1K}$ & 5714(66) & 14.0(1.9) &  & \\
${B_0^*K^*}$ & 5676(92) & 13.6(1.6) & & \\
 &  &  & 5700(81) & 16.2(2.6)\\
%
{\it Tetraquark}\\
${PA}$ & 5700(84) & 19.1(3.9) & & \\
${SV}$ & 5711(79) & 17.4(2.4) & &\\
%
\\
\hline\hline

\end{tabular*}
\label{results}
\end{table*}
\vspace*{0.6cm}
 \begin{table*}[hbt]
 	\setlength{\tabcolsep}{1.5pc}
 	\caption{LSR predictions, at NLO, for the decay constants and 
 		masses of the first radial excitation ($f_1$, $M_1$), 
 		for the molecules, tetraquark  and the predicted tetramole states. The symbol ``$\ast$'' indicates that $(BK)_1$ did not contribute to the tetramole results}
 	\label{tab:res1}
 	\begin{tabular*}{\textwidth}{@{}l@{\extracolsep{\fill}}
 			cccc}
 		\hline
 		\hline
 		Observables\, & $M_1$ (MeV) & $f_1$ (keV) 
 		&  $M_{(\mathcal T_{\mathcal M_{0,1}})_1}$ (MeV) & $f_{(\mathcal T_{\mathcal M_{0,1}})_1}$ (keV)\\
 		\hline\\
 		{\bf \boldmath $0^+$ States}\\
 		\cline{0-0}
 		{\it Molecule}\\
 		$(BK)_1$ & 6265(146) & 22.8(3.2) &  $\ast$ &  $\ast$ \\
 		$({B^*K^*})_1$ & 7494(232) & 89.2(8.4) & & \\
 		&  &  &7439(314) & 74.7(8.4)\\
 		%
 		{\it Tetraquark}\\
 		${(SS)_1}$ &7408(429) & 51.5(6.8) &  & \\
 		$({AA})_1$ & 7415(240) & 78.3(9.7) &  & \\\\
 		%
 		{\bf \boldmath $1^-$ States}\\
 		\cline{0-0}
 		{\it Molecule}\\
 		$({B_1K})_1$ & 7578(311) & 27.3(6.7) &  & \\
 		$({B_0^*K^*})_1$ & 7459(256) & 37.0(5.1) & & \\
 		&  &  & 7544(345) & 33.0(6.7)\\
 		%
 		{\it Tetraquark}\\
 		$({PA})_1$ & 7568(405) & 34.4(7.8) & & \\
 		$({SV})_1$ & 7569(388) & 32.4(7.1) & &\\
 		%
 		\\
 		\hline\hline
 		
 	\end{tabular*}
 	\label{results1}
 \end{table*}
 \vspace*{-0.6cm}

\end{document}